\documentclass[11pt]{article}%
\usepackage{amsfonts}
\usepackage{amsmath}
\usepackage{amssymb}
\usepackage{graphicx}%
\providecommand{\U}[1]{\protect\rule{.1in}{.1in}}
\newtheorem{theorem}{Theorem}
\newtheorem{acknowledgement}[theorem]{Acknowledgement}

\newtheorem{claim}[theorem]{Claim}

\begin{document}

\title{Quantum Indeterminacy and Polar Duality: a Probabilistic Approach}
\author{Maurice de Gosson\\Austrian Academy of Sciences\\Acoustics Research Institute\\1010, Vienna, AUSTRIA}
\maketitle

\begin{abstract}
We present a probabilistic argument supporting the application of polar
duality, as discussed in our previous work, to express the indeterminacy
principle of quantum mechanics. Our approach combines the properties of the
Mahler volume of a convex body with the Donoho--Stark uncertainty principle
from harmonic analysis, which characterizes the concentration of a function
and its Fourier transform. The central result demonstrates that the sum of the
probabilities of position concentration near a convex body and momentum
concentration near its polar dual is equal to one, with an error term that
diminishes rapidly as the number of degrees of freedom increases. This result
motivates the interpretation of polar duality as a kind of geometric Fourier transform.

\end{abstract}


\section{Introduction}

The term \textquotedblleft quantum indeterminacy\textquotedblright\ refers to
a fundamental concept in quantum mechanics that highlights the intrinsic
uncertainty and unpredictably in the behavior of quantum systems. There are
several different ways to express quantum indeterminacy; the simplest (from
which actually most others are derived) is that a function and its Fourier
transform cannot simultaneously sharply located. On a more sophisticated
level, it is expressed by the Heisenberg uncertainty relations (or their
refinement, the Robertson--Schr\"{o}dinger inequalities). The drawback of the
latter is that they privilege variances (and covariances) for measuring
uncertainties, which has been discussed and criticized by (Hilgevoord and
Uffink \cite{hi02,hiuf} who point out that their use for measuring the spread
is only optimal for states that are Gaussian, or close to Gaussian states. In
previous work \cite{gopol1,gopol2,gopol3,gopol4,gopolar} we have proposed a
version of quantum indeterminacy using the geometric concept of \emph{polar
duality}. We made the following physical assumption:

\begin{claim}
\textit{Let }$\mathcal{Q}$ \textit{ be quantum system centered at the origin
in phase space. If }$\mathcal{Q}$ is localized \textit{n the position
representation near a symmetric compact convex set }$X$\textit{ it cannot
simultaneously be localized in the momentum representation within a region
that is smaller than its polar dual }$X^{\hbar}$\textit{, (the polar dual
\ }$X^{\hbar}$ of $X$ \textit{the set of all }values $p$ of momentum such that
$px\leq\hbar$ for all $x$ in $X$).
\end{claim}

This a claim was motivated by reasons explained in Section 2 below; it is
however is a \emph{conjecture}, and could be verified -- or falsified -- by
experiments in the lab, or by computer simulations. That this conjecture
cannot be true in its present form is however rather obvious using the
following elementary argument: suppose we are dealing with a pure quantum
state represented by a (normalized) wavefunction $\psi$. The probability that
this state is localized in a subset $X$ of position space is
\[
\Pr(x\in X)=\int_{X}|\psi(x)|^{2}dx
\]
and the probability that it is located in the subset $X^{\hbar}$ of momentum
space is%
\[
\Pr(p\in X^{\hbar})=\int_{X^{\hbar}}|\widehat{\psi}(p)|^{2}dp
\]
where $\widehat{\psi}$ is the Fourier transform of of $\psi$. Our Claim can be
reformulated as the pair of equalities
\[
\Pr(x\in X)=\Pr(p\in X^{\hbar})=1
\]
which is not possible: since $X$ is a compact set so is $X^{\hbar}$ hence a
contradiction: a function and its Fourier transform cannot both be
simultaneously compactly supported:if $\psi$ is, then $\widehat{\psi}$ is an
analytic function, and cannot have compact support unless it is identically
zero. We will show that Claim 1 has to be replaced with the following:

\begin{claim}
\textit{Let }$\mathcal{Q}$ \textit{ be quantum system centered at the origin
in phase space and }$X$\textit{,}$X^{\hbar}$ as above\textit{. If
}$\mathcal{Q}$ is localized \textit{n the position representation near a
symmetric compact convex set }$X$\textit{ Let }$X$ and $X^{\hbar}$\textit{be
as above}. We have $\Pr(x\in X)+\Pr(p\in X^{\hbar})=1$ up to a an error
tending rapidly to $0$ as $n\rightarrow\infty$.
\end{claim}

We will see that the proof of this claim uses non-trivial methods borrowed
from both geometry (the Blaschke--Santal\'{o} inequality) and analysis (the
Donoho--Stark uncertainty principle) as will be discussed in Section 4. 

\section{Proprieties of Polar Duality}

Let $X$ be anon-empty convex compact set in position space $\mathbb{R}_{x}%
^{n}$. We assume that $X$ is symmetric (hence it contains $0$). By definition
\cite{Vershynin}, the polar dual $X^{\hbar}$ is the set of all $p=(p_{1}%
,...,p_{n})$ in momentum space $\mathbb{R}_{p}^{n}$ such that%
\[
px=p_{1}x_{1}+\cdot\cdot\cdot+p_{n}x_{n}\leq\hbar.
\]
Notice that the larger $X$ is, the smaller $X^{\hbar}$ is, and vice-versa.

We briefly recall the basic properties of polar duality as exposed for
instance in \cite{gopol2}. The assignment $X\longrightarrow X^{\hbar}$ is
reflexive: $(X^{\hbar})^{\hbar}=X$, anti-monotone: if $X\subset Y$ then
$Y^{\hbar}\subset X^{\hbar}$, and has the covariance property under linear
transformations: if $\det A\neq0$ then
\begin{equation}
(AX)^{\hbar}=(A^{T})^{-1}X^{\hbar}. \label{scale}%
\end{equation}
Let $B_{X}^{n}(R)$ (\textit{resp}. $B_{P}^{n}(R)$) be the ball defined by
$|x|\leq R$ in $\mathbb{R}_{p}^{n}$\ (\textit{resp}. $|p|\leq R\}$ in
$\mathbb{R}_{p}^{n}$). Then $B_{X}^{n}(R)^{\hbar}=B_{P}^{n}(\hbar/R)$ and
hence, particular,
\begin{equation}
B_{X}^{n}(\sqrt{\hbar})^{\hbar}=B_{P}^{n}(\sqrt{\hbar}). \label{bhh}%
\end{equation}
Let $A$ be a real invertible and symmetric $n\times n$ matrix and $R>0$. It
follows from (\ref{scale}) that the polar dual of the ellipsoid defined by
$Ax\cdot x\leq R^{2}$ is given by $A^{-1}p\cdot p\leq(\hbar/R)^{2}\}$ and
hence%
\begin{equation}
\{x:Ax\cdot x\leq\hbar\}^{\hbar}=\{p:A^{-1}p\cdot p\leq\hbar\}.
\label{dualellh}%
\end{equation}

The reason why we have introduced the notion of polar duality in the study of
quantum indeterminacy comes from the following observation: Assume that $X$ is
an ellipsoid, the so is $X^{\hbar}$, in view of (\ref{dualellh}), and we have proved

\begin{claim}
The phase space cell $X\times$ $X^{\hbar}$ contains a unique quantum blob.
\end{claim}

Recall \cite{blobs,golu09} that a \textquotedblleft quantum
blob\textquotedblright\ is the image of the phase space ball $B^{2n}%
(\sqrt{\hbar})$ by a linear symplectic transformation $S\in\operatorname*{Sp}%
(n)$.This is easily seen as follows: the ellipsoid $X:Ax\cdot x\leq\hbar$ is
the image of the ball $B_{X}^{n}(\sqrt{\hbar})$ by the linear mapping
\ $A^{-1/2}$ while the ellipsoid $X^{\hbar}:A^{-1}p\cdot p\leq\hbar$ is that
of $B_{P}^{n}(\sqrt{\hbar})$ by $A^{1/2}$. It follows that the cell $X\times$
$X^{\hbar}$ is the image of the product $B_{X}^{n}(\sqrt{\hbar})\times
B_{P}^{n}(\sqrt{\hbar})$ by the symplectic mapping $S=%
\begin{pmatrix}
A^{-1/2} & 0\\
A^{-1/2} & A^{1/2}%
\end{pmatrix}
$. Now the unique largest ellipsoid (John ellipsoid, see \cite{Ball})
contained in $B_{X}^{n}(\sqrt{\hbar})\times B_{P}^{n}(\sqrt{\hbar})$ is
$B^{2n}(\sqrt{\hbar})$ hence the John ellipsoid of $X\times$ $X^{\hbar}$ is
$S(B^{2n}(\sqrt{\hbar})),$ which is a quantum blob.

\ Quantum blobs represents the smallest unit of phase space compatible with
the uncertainty principle. They is defined in the context of the
Robertson-Schr\"{o}dinger uncertainty relation and are characterized by its
invariance under symplectic transformations. To every quantum blob one
associates in a canonical way a generalized coherent state. For instance , to
the ball $B^{2n}(\sqrt{\hbar})$ is associated the $n$-dimensional coherent
state $\phi_{0}(x)=(\pi\hbar)^{-n/4}e^{-x^{2}/2\hslash}$. We have detailed
these properties in our paper \cite{golu09} with Luef.

\section{Mahler's Volume and the Blaschke--Santal\'{o} Inequality}

A remarkable property of polar duality, the Blaschke--Santal\'{o} inequality:
says that if $X$ is a symmetric convex body; then the Mahler volume $v(X)$,
defined by%
\[
v(X)=(\operatorname*{Vol}X)(\operatorname*{Vol}X^{\hbar})
\]
satisfies the inequality%
\begin{equation}
v(X)\leq(\operatorname*{Vol}\nolimits_{n}(B^{n}(\sqrt{\hbar}))^{2}%
\label{santalo1}%
\end{equation}
that is,
\begin{equation}
v(X)=(\operatorname*{Vol}X)(\operatorname*{Vol}X^{\hbar})\leq\frac{(\pi
\hbar)^{n}}{\Gamma(\frac{n}{2}+1)^{2}}\label{BS0}%
\end{equation}
where $\operatorname*{Vol}\nolimits_{n}$ is the standard Lebesgue measure on
$\mathbb{R}^{n}$, and equality is attained if and only if $X\subset
\mathbb{R}_{x}^{n}$ is an ellipsoid centered at the origin. The Mahler has
conjectured that lower bound is%
\begin{equation}
\upsilon(X)\geq\frac{(4\hbar)^{n}}{n!}\label{volvo3}%
\end{equation}
but this claim has so far resisted to all proof attempts The best know result
is the following, due to Kuperberg \cite{Kuper}, who has shown that
\begin{equation}
\upsilon(X)\geq\frac{(\pi\hbar)^{n}}{4^{n}n!}.\label{kuper}%
\end{equation}
Summarizing, we have the bounds%
\begin{equation}
\frac{(\pi\hbar)^{n}}{4^{n}n!}\leq\upsilon(X)\leq\frac{(\pi\hbar)^{n}}%
{\Gamma(\frac{n}{2}+1)^{2}}\label{bounds}%
\end{equation}
\ (see \cite{gopolar} for a discussion of other partial results).

The Mahler volume has the intuitive interpretation as being a measure of
\textquotedblleft roundness\textquotedblright: its largest value is taken by
balls (or ellipsoids), and its smallest value (the bound (\ref{volvo3})) is
indeed attained by any $n$-parallelepiped%
\begin{equation}
X=[-\sqrt{2\sigma_{x_{1}x_{1}}},\sqrt{2\sigma_{x_{1}x_{1}}}]\times\cdot
\cdot\cdot\times\lbrack-\sqrt{2\sigma_{x_{n}x_{n}}},\sqrt{2\sigma_{x_{n}x_{n}%
}}].\label{interval}%
\end{equation}
This is related to the covariances of the tensor product $\psi=\phi_{1}%
\otimes\cdot\cdot\cdot\otimes\phi_{n}$ of standard one-dimensional Gaussians
$\phi_{j}(x)=(\pi\hbar)^{-1/4}e^{-x_{j}^{2}/2\hbar}$; the function $\psi$ is a
minimal uncertainty quantum state in the sense that it reduces the Heisenberg
inequalities to equalities. We suggest that such quantum considerations might
lead to proof of Mahler's conjecture.

\subsection{The Donoho--Stark inequality}

A remarkable result is the Donoho--Stark inequality, related to uncertainty
principles, particularly in signal processing and harmonic analysis
\cite{gro}. This inequality addresses the relationship between the sparsity of
a signal and its representation in dual domains, such as time and frequency.
It can be stated as follows: let $\psi\in L^{2}(\mathbb{R}^{n})$ and
\begin{equation}
\widehat{\psi}(p)=\left(  \tfrac{1}{2\pi\hbar}\right)  ^{n/2}\int%
_{\mathbb{R}^{n}}e^{-ipx/\hbar}\psi(x)dx \label{FT}%
\end{equation}
be its $\hbar$-Fourier transform; we assume that $||\psi||_{L^{2}%
}=||\widehat{\psi}||_{L^{2}}=1$. Donoho and Stark proved in \cite{DS} that if
$X\subset\mathbb{R}^{n}$ and $P\subset(\mathbb{R}^{n})^{\ast}$ are measurable
sets such that
\begin{equation}
\int_{X}|\psi(x)|^{2}dx\geq1-\varepsilon\text{ },\text{ }\int_{P}%
|\widehat{\psi}(p)|^{2}dp\geq1-\eta\label{DS1}%
\end{equation}
where $\varepsilon$ and $\eta$ are non-negative numbers such that
$0\leq\varepsilon+\eta<1$, then the volumes of $X$ and $P$ cannot be
simultaneously arbitrarily small, in fact, we must have
\begin{equation}
(\operatorname*{Vol}X)(\operatorname*{Vol}P)\geq(2\pi\hbar)^{n}(1-\varepsilon
-\eta)^{2} \label{DS2}%
\end{equation}
(see the article \cite{Boggiatto} by Boggiatto \textit{et al}. for a detailed
discussion and improvements of the inequality (\ref{DS2}).

\section{The Main Result: Discussion and Interpretation}

Suppose now we take $P=X^{\hbar}$ in the Donoho-Stark inequality (\ref{DS1}).
Combining the inequalities (\ref{DS2}) and (\ref{BS0}) then yields the double
inequality%
\begin{equation}
(2\pi\hbar)^{n}(1-\varepsilon-\eta)^{2}\leq(\operatorname*{Vol}%
X)(\operatorname*{Vol}X^{\hbar})\leq\frac{(\pi\hbar)^{n}}{\Gamma(\frac{n}%
{2}+1)^{2}}\label{DSBS}%
\end{equation}
which implies that we must have
\begin{equation}
0\leq1-\varepsilon-\eta\leq\delta(n)\text{ },\text{ \ }\delta(n)=\frac
{1}{2^{n/2}\Gamma(\frac{n}{2}+1)}%
\end{equation}
that is, equivalently,%
\begin{equation}
1-\delta(n)\leq\varepsilon+\eta\leq1.\label{epsilon}%
\end{equation}
Adding the probabilities%
\begin{align*}
\Pr(x &  \in X)=\int_{X}|\psi(x)|^{2}dx=1-\varepsilon\\
\Pr(p &  \in X^{\hbar})=\int_{X^{\hbar}}|\widehat{\psi}(p)|^{2}dp=1-\eta
\end{align*}
we have, taking the inequalities  (\ref{epsilon}) into account,%
\begin{equation}
1\leq\Pr(x\in X)+\Pr(p\in X^{\hbar})\leq1+\delta(n)\label{51}%
\end{equation}
that is%
\begin{gather}
\Pr(x\in X)+\Pr(p\in X^{\hbar})=1+\Delta(n)\\
0\leq\Delta(n)\leq\delta(n).
\end{gather}
Thus, $\Pr(x\in X)+\Pr(p\in X^{\hbar})$ is very close to one, even for
relatively small values of $n$. In fact, using Stirling's formula
\[
\Gamma(m+1)\overset{m\rightarrow\infty}{\backsim}\sqrt{2\pi m}(m/e)^{m}%
\]
we have the estimate%
\begin{equation}
\Pr(x\in X)+\Pr(p\in X^{\hbar})\overset{nm\rightarrow\infty}{\backsim}%
1+\sqrt{\frac{\pi n}{2^{n}}}\left(  \frac{n}{2e}\right)  ^{n/2}\label{52}%
\end{equation}
for large values of the dominion $n$. These formulas clearly show the
\textquotedblleft trade-off\textquotedblright\ between the concentration of
position values and momentum values. The Mahler volume being a measure of the
\textquotedblleft roundness\textquotedblright\ \ of $X$, the estimates above
are optimal when $X$ is nearly an ellipse. In the case of an $n$%
-parallelepiped, we get thew sharper estimate
\begin{gather}
\Pr(x\in X)+\Pr(p\in X^{\hbar})=1+\Delta_{\min}(n)\\
0\leq\Delta_{\min}(n)\leq\upsilon(X)\geq\frac{(\pi\hbar)^{n}}{4^{n}n!}%
\end{gather}
where we have taken into account the Kuperberg lower bound (\ref{kuper}).

The results above can be interpret as a trade-off between a wave function and
its Fourier transform. Assume for instance that we have $\varepsilon
\thickapprox1$. Then $\Pr(x\in X)\thickapprox0$  and $\Pr(p\in X^{\hbar
})\thickapprox1$. This can be interbred by saying that if the quantum system
cannot be localized \textquotedblleft anywhere\textquotedblright\ in position
space then it is most certainly localized near the origin in momentum space
(cf. the Heisenberg inequality). \ If, on the other hand, the system under
consideration has around $50\%$ probability being localized inside $X$ in
position space then it also has a $50\%$ probability of being localized in
$X^{\hbar}$.

Here is a simple illustration: choose $n=1$, $X=B^{1}(\sqrt{\hbar}%
)=[-\sqrt{\hbar}),\sqrt{\hbar})]$ and $\psi=\phi_{0}$ (the fiducial coherent
state: $\phi_{0}(x)=(\pi\hbar)^{-1/4}e^{-x^{2}/2\hbar}$. Then $X=X^{\hbar}$,
\ $\widehat{\phi}_{0}=\phi_{0}$, and (\ref{52}) yields
\[
\Pr(x\in X)=\Pr(p\in X^{\hbar})\thickapprox\frac{1}{2}(1+\delta(1))=0.813
\]
which is to be compared with the tabulated value $\Pr(x\in X)\thickapprox
0.8427$.

Our result is refinement of a previous work of ours where we discussed Hardy's
uncertainty principle \cite{Hardy}. We showed in \cite{ACHA} that if $\psi$ is
a non-zero square integrable function satisfying estimates%
\[
|\psi(x)|\leq C_{A}e^{-Ax^{2}/2\hbar}\text{ \ },\text{ \ }|\widehat{\psi
}(p)|\leq C_{B}e^{-Bx^{2}/2\hbar}%
\]
where $C_{A},C_{B}\geq$ and $A,B$ are positive definite symmetric matrices,
then the ellipsoids%
\[
X_{A}:\{x:Ax^{2}\leq\hbar\}\text{ \ , \ }P_{B}:\{x:Bp^{2}\leq\hbar\}
\]
satisfy $X_{A}^{\hbar}\subset P_{B}$, with equality $X_{A}^{\hbar}=P_{B}$ (
corresponding to the case $A=B^{-1}$.) if and only if $\psi(x)=Ce^{-Ax^{2}%
/2\hbar}$ foe some complex number $C\neq0$,

\begin{acknowledgement}
This work has been financed by the Austrian Research Foundation FWF (Grant
number PAT 2056623). It was done during a stay of the author at the Acoustics
Research Institute group at the Austrian Academy of Sciences. We thank Peter
Balazs for his hospitality.
\end{acknowledgement}

MauriceAlexis.deGossondeVarennes@oeaw.ac.at

maurice.de.gosson@univie.ac.at

\end{document}